\documentclass[namedreferences]{solarphysics}

\usepackage{etoolbox}
\usepackage[hyperref,optionalrh,showbiblabels]{spr-sola-addons} 
\usepackage{graphicx}        
\usepackage{color}           

\usepackage[english]{babel}
\usepackage[dvipsnames,svgnames,x11names]{xcolor}

\renewcommand{\vec}[1]{{\mathbfit #1}}

\newcommand{\curl}{ {\bf \nabla} \times}

\newcommand{\bb}{\vec B}

\newcommand{\adv}{    {\it Adv. Space Res.}}

\newcommand{\apj}{    {\it Astrophys. J.}}
\newcommand{\apjl}{   {\it Astrophys. J. Lett.}}

\newcommand{\solphys}{{\it Solar Phys.}}
 
\newcommand{\ssr}{    {\it Space Sci. Rev.}} 
\chardef\us=`\_

\begin{document}

\begin{article}
\begin{opening}

\title{Inflows towards Bipolar Magnetic Active Regions and Their Nonlinear Impact on a Three-Dimensional Babcock-Leighton Solar Dynamo Model\\ {\it Solar Physics}}

\author[addressref={aff1,aff2},corref,email={kinntt2000@mail.com}]{\inits{Kinfe}\fnm{Kinfe}~\lnm{Teweldebirhan}}
\author[addressref=aff3,email={mark.miesch@noaa.gov}]{\inits{Mark}\fnm{Mark}~\lnm{Miesch}}
\author[addressref=aff2,email={sgibson@ucar.edu}]{\inits{Sarah}\fnm{Sarah}~\lnm{Gibson}}

\address[id=aff1]{Physics Department, College of Natural and Computational Sciences, Aksum University, Aksum, Tigray}
\address[id=aff2]{High Altitude Observatory, National Center for Atmospheric Research, 3080 Center Green Dr., Boulder, CO 80301}
\address[id=aff3]{Cooperative Institute for Research in Environmental Sciences, CU Boulder, NOAA Space Weather Prediction Center, NOAA David Skaggs Research Center, 325 Broadway, Boulder, CO 30305}

\runningauthor{Teweldebirhan et al. }
\runningtitle{BMR Inflows and the Solar Dynamo}

\begin{abstract}

The changing magnetic fields of the Sun are generated and maintained by a solar dynamo, the exact nature of which remains an unsolved fundamental problem in solar physics. Our objective in this paper is to investigate the role and impact of converging flows toward Bipolar Magnetic Regions (BMR inflows) on the Sun's global solar dynamo. These flows are large-scale physical phenomena that have been observed and so should be included in any comprehensive solar dynamo model. We have augmented the Surface ﬂux Transport And Babcock–LEighton (STABLE) dynamo model to study the nonlinear feedback effect of BMR inflows with magnitudes varying with surface magnetic fields. This fully-3D realistic dynamo model produces the sunspot butterfly diagram and allows a study of the relative roles of dynamo saturation mechanisms such as tilt-angle quenching and BMR inflows. The results of our STABLE simulations show that magnetic field dependent BMR inflows significantly affect the evolution of the BMRs themselves and result in a reduced buildup of the global poloidal field due to local flux cancellation within the BMRs, to an extent that is sufficient to saturate the dynamo. As a consequence, for the first time, we have achieved fully 3D solar dynamo solutions in which BMR inflows alone regulate the amplitudes and periods of the magnetic cycles. 

\end{abstract}
\end{opening}

\section{Introduction}
\label{S-Introduction} 

The Sun's magnetic field is generated by a dynamo process in the convection zone of the solar interior, which is based on the nonlinear interaction between the velocity field and the magnetic field of the solar plasma. This nonlinear interaction is mathematically described by the well-known magnetohydrodynamical (MHD) equations, and in particular, the evolution of the magnetic field, $\vec{B}$, in response to the velocity field, $\vec{v}$  is given by the induction equation. as described in our previous studies \citep{miesc14, miesc16}. As demonstrated in \citet{bhowmik23}, these equations allow prediction of the strength of the solar cycle and cycle-to-cycle fluctuations, which determine the level of solar activity, including flares and coronal mass ejections and associated space weather.

The solar dynamo process, which is the generation and maintenance of the Sun's magnetic field, involves a complex dynamic interaction of differential rotation, meridional circulation, turbulent diffusion and other convection flows. The deep-seated toroidal magnetic fields generated in the convection zone manifest in the cyclic appearance of magnetized sunspots on the solar surface (photosphere) as these magnetic fields rise to the surface through a process, known as magnetic flux emergence, that is considered by many to be central to the dynamo process \citep{fan09, camer15, miesc16, chrb20}. In this paper, we will refer to these emerged sunspots in general as Bipolar Magnetic Regions (BMRs).

A variety of helioseismic observations have detected and analyzed spatially extended flows converging into solar active regions near the surface (i.e., BMR inflows; - see Section~\ref{sec-con-flow}).
These BMR inflows may hold the solution to one of the crucial issues facing dynamo models, i.e., establishing what stops the magnetic field from increasing without bound, or in other words, identifying a dynamo saturation mechanism. The MHD induction equation in kinematic dynamo theory is linear, admitting solutions that either grow or decay exponentially. Thus, some nonlinearity is necessary to result in the regular magnetic cycles that are observed. 

BMR inflows may provide a nonlinear feedback mechanism limiting the amplitude of a Babcock-Leighton type solar dynamo model (Section~\ref{S-kinematic}). In this framework, surface flux transport determines the variation of the cycle strength by modulating the buildup of polar fields \citep{jiang14, upton14}. BMR inflows have an essential role in the concentration and dispersal of surface magnetic flux and may be responsible for the occurrence of magnetic reconnection by bringing together opposite magnetic polarities, leading to flux cancellation and decreasing the amount of magnetic flux in solar active regions. This magnetic flux cancellation suppresses or interrupts the solar dynamo process to some extent, thus inhibiting further amplification of the magnetic field. Therefore, flux cancellation due to BMR inflows can play a significant role in the saturation of the solar dynamo. 

In surface flux transport models by \citep{jiang10, camers10} the mean poleward meridional flow varies due to the surface BMR inflow. This can affect the surface magnetic field and polar field resulting from the latitudinal transport of surface flux.
Surface flux transport models (\citealp{camer12, belda17}) have found that axisymmetric latitudinal inflow into active region belts can modulate the solar cycle within the Babcock-Leighton framework in a manner that is consistent with BMR inflows playing a significant role in the saturation of the global dynamo. More recently, \citet{nagy20} also found enhanced dynamo saturation in a $2\times 2D$ Babcock–Leighton solar cycle model which included an explicit nonlinear backreaction mechanism where the width and speed of an axisymmetric inflow into the active-region belt was proportional to the magnitude of the emerging flux. However, when they removed an additional saturating term (tilt-angle quenching; see Section~\ref{sec-spotmaker}), they found that BMR inflow on its own was not enough to saturate the dynamo.

None of the models employed in addressing this question so far have been fully 3D, nor used BMR inflows as a saturation mechanism alone.
In this paper, we introduce BMR inflow into the fully 3D STABLE (Surface ﬂux Transport And Babcock–LEighton) solar dynamo model \citep{miesc14, miesc16, hazra16, karak17}. This allows us to investigate the extent to which the observed near-surface inflows towards active regions can provide a nonlinear feedback mechanism that limits the amplitude of a Babcock-Leighton-type solar dynamo and determines the variation of the cycle strength. 

In Section \ref{S-kinematic}, we describe the formulation of the STABLE 3D solar dynamo model and the imposed ﬂow ﬁelds in the context of kinematic induction as well as the implemention of the BMR emergence into STABLE. In Section \ref{sec-con-flow}, we describe the detailed mathematical formulation of BMR inflows and discuss the implications for meridional flow. In Section \ref{sec:inflowpolarfield}, we provide a parameter study of the impacts of BMR inflow on polar field buildup for the illustrative cases of one and two (one in each hemisphere) BMRs.  In Section \ref{sec:solaractivity} we demonstrate the impacts of magnetic field dependent BMR inflows for sustaining a solar dynamo simulations over many magnetic solar cycles and compare STABLE results for both the tilt-angle-quenching and BMR-inflow dynamo saturation mechanisms.  We present our conclusions in Section \ref{S-Conclusion}.

\section{The STABLE solar dynamo model — kinematic induction}
\label{S-kinematic}

The magnetic field of the Sun is generated by dynamo action in its conducting interior leading to cyclic evolution of solar magnetic features. A hydromagnetic dynamo is a process by which the magnetic field in an electrically conducting fluid is maintained against Ohmic dissipation 
through a nonlinear interaction between the velocity field and the magnetic field of the solar plasma. 
This nonlinear interaction is mathematically described by the well-known magnetohydrodynamical (MHD) equations. The evolution of the magnetic field, $\vec{B}$, in response to the velocity field, $\vec{v}$ and turbulent diffusion $\eta_t$, is given by the induction equation:
\begin{equation}\label{eq:indy}
\frac{\partial \bb}{\partial t} = \curl \left(\vec{v} \times \bb - \eta_t \curl \bb\right) 
\end{equation}

Thus, the dynamo problem consists of producing a flow field $\vec{v}$ that has inductive properties capable of sustaining $\vec{B}$ against turbulent diffusion. The toroidal magnetic field is generated from differential rotation operating on a preexisting poloidal component, and in flux-transport dynamos, the poloidal field is generated from a preexisting toroidal component that emerges through the solar surface (see \citet{chrb20, camer23} and references therein). 

The STABLE dynamo model solves the kinematic MHD induction equation as first reported in \citet{miesc14} and later in more detail by \citet{miesc16}, \citet{hazra16} and \citet{karak17}. The Anelastic Spherical Harmonic (ASH) code \citep{brun04, miesc05, miesc09, brun10b} serves as the  dynamical core for the STABLE model. 
STABLE is a 3D Babcock-Leighton/Flux Transport dynamo model in which the source of the poloidal field is determined from the explicit emergence, distortion, and dispersal of BMRs by differential rotation, meridional circulation and now inflow into the BMRs (see Figure \ref{F-mag-Br}). 

The velocity field $\vec{v}$ in STABLE is inferred from observations and is defined by the total sum of flow fields including differential rotation $\textbf{v}_{DR}$, meridional circulation $\textbf{v}_{mc}$, 
as well as the magnetic field dependent BMR inflows $\textbf{v}_{in}$:

\begin{equation}\label{eq:tflows}
\textbf{v}(r, \theta, \phi, t) = \textbf{v}_{DR}(r, \theta) + \textbf{v}_{mc}(r, \theta) + \textbf{v}_{in} (r, \theta, \phi, t)
\end{equation}

Here $\theta$ and $\phi$ are heliospheric colatitude and longitude respectively.  The components $v_{in\phi}$, $v_{in\theta}$ and $v_{inr}$ of the inflows are the longitudinal, latitudinal and radial components in spherical coordinates (see Section~\ref{sec-con-flow}).

We now provide details about the formulation of the STABLE model through the imposed
 mean flows (i.e., $\textbf{v}_{DR}$ and $\textbf{v}_{mc}$) and turbulent diffusivity $\eta_t$ (Section~\ref{s-flows}, and describe the method for emerging BMRs (SpotMaker; Section~\ref{sec-spotmaker}). We will return to the discussion and inclusion of $\textbf{v}_{in}$ in Section \ref{sec-con-flow}.

\subsection{Imposed Mean Flows and Turbulent Diffusivity}
\label{s-flows}

The meridional circulation, differential rotation, and turbulent diffusion are essential dynamo model ingredients that govern STABLE \citep{miesc16,karak17}.  The mean meridional circulation is an axisymmetric flow (averaged over longitude) with a single cell assumed per hemisphere directed poleward at the surface and equatorward near the base of the convection zone.  Here we use the same profile used in \citet{karak17}, which is similar to many previous 2.5D (axisymmetric) and 3D Babock-Leighton models inspired by observations.   The maximum flow speed is about 20 m s$^{-1}$ at the surface and about 2 m s$^{-2}$ near the base of the convection zone. The mean differential rotation profile, $\Omega(r,\theta)$ is also formulated as in previous STABLE models and other Babcock-Leighton models.  For further details see \citet{miesc16} (Equation 3 and Figure 1).  
In order to build on previous work and to easily compare with previous results, we also use the same turbulent diffusion and turbulent pumping profile as in the diffusion-dominated model described by \citet{karak17}.  Together with the meridional circulation profile, this gives a more realistic poleward flux transport compared to that shown in \citet{miesc16}.

\subsection{Magnetic Field Dependent BMR Emergence: SpotMaker Algorithm}
\label{sec-spotmaker}

The SpotMaker spot deposition algorithm is unique to the STABLE solar dynamo model. As in previous studies \citep{miesc14, miesc16, hazra16, karak17}, SpotMaker places tilted BMRs on the solar surface in response to the dynamo-generated toroidal magnetic field. The subsequent decay and dispersal of these BMRs due to differential rotation, meridional circulation, and turbulent diffusion generates poloidal field as described by \citet{babco61} and \citet{leigh64}.  The production of the poloidal field depends on the amount of magnetic flux in BMRs,  the frequency of BMR eruptions, and the tilt angles of BMRs.   Toroidal field is generated from this poloidal field through the mechanism of differential rotation \citep{miesc16}. The STABLE solar dynamo model constructed in this fashion is sufficient to establish and maintain magnetic cycles.

\begin{figure}    
\centerline{\includegraphics[height=7cm,width=1.0\textwidth,clip=]{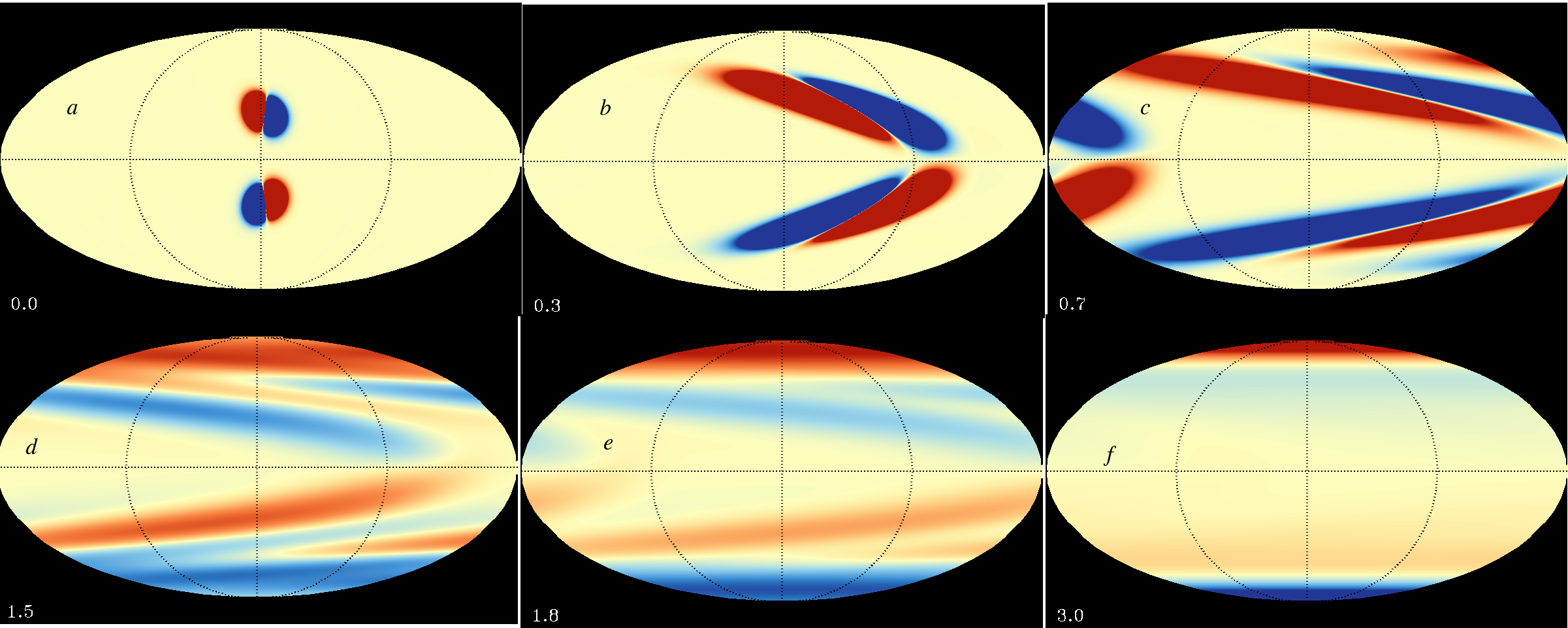}}

\caption{The time evolution (a-f) of surface radial magnetic field of the sun with a single pair of BMRs. The BMRs are placed one each in the northern and southern hemisphere and are initially located at $\pm 25^\circ$, with opposite polarity ordering in each hemisphere, as per Hale's polarity laws. The surface field evolves in response to diffusion and advective transport by differential rotation and a poleward meridional flow, for $(a)$ 0.0 years, $(b)$ 0.3 years, $(c)$ 0.7 years, $(d)$ 1.5 years, $(e)$ 1.8 years and (f) 3.0 years. Here the red color shows the outward-directed radial field and the blue color represents the inward-directed radial field.  }
   \label{F-mag-Br}
   \end{figure}
   
The emergence rate of BMRs depends on the integrated toroidal flux near the base of the convection zone:
\begin{equation}\label{eq:spotoma}
\hat{\textbf{B}}_\phi(\theta, \phi, t) = \int^{r_b}_{r_a}h(r)\textbf{B}_\phi(r, \theta, \phi, t)dr  ~~~.
\end{equation}
where, $r_a$ = 0.715R, $r_b$ = 0.73R, and $h_(r)$ as a normalization factor. For further details see \citet{miesc14, miesc16, karak17}. Whether or not a spot is placed at the solar surface depends on whether $\textbf{B}_\phi$ below it is greater than a threshold value assumed by the model; note this threshold toroidal field for producing new BMRs increases exponentially with latitude as described by \citet{karak17} (their equation 6) in order to suppress unrealistic spot emergence at high latitudes.

The timing of spot placement is done randomly based on a lognormal probability density distribution with a mean ($\tau_p$) and mode ($\tau_s$) given by \citet{karak17}:
\begin{equation}\label{eqch5:mean}
 \tau_p = \frac{2.2 ~\mbox{days}}{1 + (B_b/B_\tau)^2} ~~~, ~~~ \tau_s = \frac{20 ~\mbox{days}}{1 + (B_b/B_\tau)^2} ~~~.
\end{equation}
where $B_b$ is the mean value of $\hat{B}$ in equation (\ref{eq:spotoma}) at $\pm$ 15$^\circ$ latitude and averaged over longitude.  The value of $B_\tau$ is calibrated to 400 G so the number of BMRs that is produced is comparable to solar observations.

The flux distribution of BMRs is also chosen randomly from a lognormal distribution calibrated to match solar observations \citep{karak17}.  Hale's polarity law is not explicitly imposed.  Rather, the sign of the leading and trailing polarity is determined from the dynamo-generated toroidal field near the base of the convection zone.

The 3D structure of the BMRs that are produced by SpotMaker is described in detail by \citet{miesc16}.  The surface field in each BMR is localized and has an equal amount of positive and negative flux (see, e.g., Figure~\ref{F-mag-Br} top left).  The orientation of each BMR follows Joy's law with an optional tilt angle quenching term that limits the degree of tilt ($\delta$) as toroidal magnetic field becomes large:
\begin{equation}\label{tiltangle_sat}
\delta = \frac{\delta_o \cos\theta}{1 + (\vec{B}_{\phi}(\theta, \phi, t)/B_{sat})} ~~~. 
 \end{equation}
Where $\delta_o~ = ~35^o$ and $B_{sat} = 1 \times 10^5 G$.  This tilt-angle quenching was used by \citet{karak17} as a dynamo saturation mechanism.  In this paper, we replace tilt-angle quenching with BMR inflows as the saturation mechanism (Section~\ref{sec:solaractivity}).  So, in our BMR-inflow simulations, we omit the denominator in Equation \ref{tiltangle_sat}.  The tilt is then just given by Joy's law.

After defining the surface magnetic field, we extrapolate the field downward to 0.9R using a poloidal magnetic potential to ensure that the 3D magnetic field in each BMR is divergenceless, thus confining magnetic flux in each BMR to near the surface, above 0.9R (see Figure~2 of \citet{miesc16}). 
  
\section{Converging Flows Towards the Bipolar Magnetic Regions (BMR Inflows)}
 \label{sec-con-flow}

Helioseismic studies have reported extended horizontal flows around active regions with speeds of tens of m/s and flows extending out to tens of degrees from the centers of active regions \citep{gizon01, haber02, spruit03, haber03, haber04, zhao04, derosa06, gizon08, Švanda08, hindman09, gizon10, kosov16, lptien17,braun19, gott21, mahajan23}. The flows converge near the surface and diverge below, implying a three-dimensional circulation (see Figure 11 of \citet{hindman09}). 
These converging flows possibly last (at least) as long as the magnetic regions themselves, and the flows 
affect larger circulation patterns, particularly the meridional 
and zonal flow components of global circulation \citep{braun19}. 

\begin{figure}    
\centerline{\hspace*{0.015\textwidth}
               \includegraphics[width=0.60\textwidth,clip=]{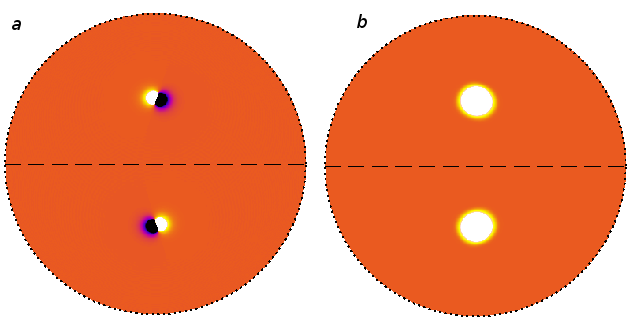}
               \hspace*{-0.02\textwidth}
             
               \includegraphics[width=0.35\textwidth,clip=]{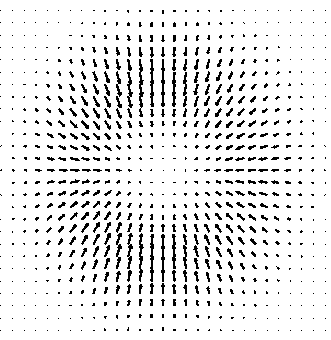}
              }
     \vspace{-0.35\textwidth}   
     \centerline{\Large \bf     
      \hspace{0.65\textwidth}  \color{black}{c}
         \hfill}
     \vspace{0.32\textwidth}    
              
\caption{STABLE creates BMR surface inflows by modeling a horizontal flow from the smoothed, unsigned radial magnetic field of the BMR. 
 ($a$) Radial magnetic field $B_r$ on the surface of the  sun with a single BMR in both the northern and southern  hemispheres. Here, white shows outward-directed field and blue shows inward-directed field. ($b$) The same, except using the absolute value of $|B_r|$ and smoothing it as described in the text. ($c$) The BMR inflow velocity field is then calculated directly from the smoothed, unsigned magnetic field in ($b$) (Equation~\ref{hvs}).}
 \label{F-Br-smooth}
   \end{figure}

It has been argued that the magnitudes of BMR inflows depend on the magnetic flux of BMRs at the solar surface, driven by the enhanced radiative cooling in the active regions \citep{spruit03, derosa06, gizon08,camers10,camer12}. We therefore model the magnetic field dependent BMR inflows at the solar surface as:

\begin{equation}\label{hvs}
\textbf{v}_{{in}_s}(\theta, \phi, t) = c \nabla_h\left(\frac{\sin \theta}{\sin 30^\circ}\frac{|B_r|_{sm}}{B_n}\right) 
\end{equation}

\begin{figure}  
   \centerline{\includegraphics[width=1.0\textwidth,clip=]{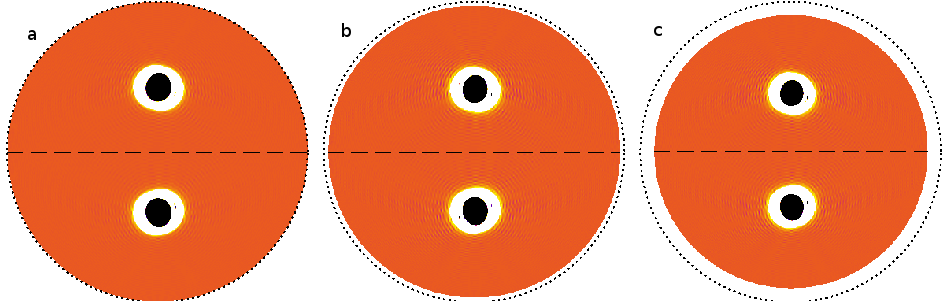}
              }
\caption{The radial flow field, with black/white showing downward draft/upward flow of mass flux. ($a$) The flow at the surface r=R; ($b$) at a radius of 0.96R; and ($c$) at a radius of 0.92R. The downward flow at the center of the BMR surface inflow is a consequence of the enhanced cooling in the BMRs.}
   \label{F-vr-inflow}
   \end{figure}

\begin{figure}
   \centerline{\includegraphics[width=1.0\textwidth,clip=]{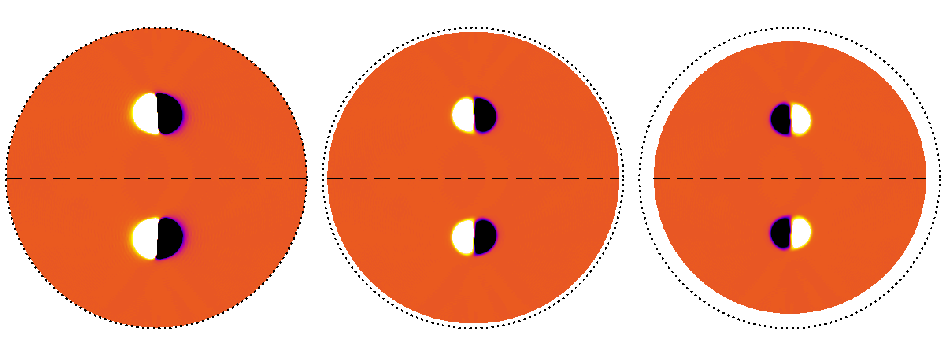}
              }
\caption{As in Figure~\ref{F-vr-inflow}, but showing the longitudinal flow field, which reverses direction at greater depth (right panel), illustrating the three-dimensional circulation of convergence near the surface and divergence below.}
   \label{F-vphi-inflow}
   \end{figure}

Here, $\nabla_h$ is the horizontal gradient operator and $c$ is an adjustable constant with units of velocity that we will specify to control the speed of the inflow.  The normalization factor $B_n$ is calibrated so that the amplitude of the inflow velocity is comparable to $c$.  
The term $\frac{\sin \theta}{\sin 30^\circ}$ suppresses unrealistically strong flow perturbations
 that would otherwise occur at high latitudes from the gradient of the polar fields.  Figure \ref{F-Br-smooth}-$a$ shows snapshots of radial magnetic field, $B_r$, on the solar surface. Figure \ref{F-Br-smooth}-b is
$|B_r|_{sm}$, which is the smoothed absolute value of the surface magnetic field in the BMR.  Smoothing ensures that the inflow occurs around the periphery of each BMR as shown in Figure \ref{F-Br-smooth}-c, which promotes flux cancellation \citep{camer12}.  The smoothing shown in Figures \ref{F-Br-smooth}--\ref{F-vphi-inflow} is done using a circular tent filter with a radius of 15 degrees. 

Since our model is 3D  and nonaxisymmetric, we will consider all components of BMR inflows; these are latitudinal, longitudinal and radial components.  These must satisfy the continuity equation, so mass is conserved and the mass flux is divergenceless.

\begin{eqnarray}\label{continuity}
\frac{1}{r^2} \frac{\partial(r^2\rho v_r)}{\partial r} = - \nabla_h\cdot(\rho \textbf{v}_h)
\end{eqnarray}
where $\nabla_h \cdot$ is the horizontal divergence and $\textbf{v}_h = \textbf{v}_\theta + \textbf{v}_\phi $ horizontal velocity respectively.

We satisfy equation  (\ref{continuity}) exactly by expressing the mass flux in terms of a poloidal (W) streamfunction as follows: 

\begin{equation}\label{polotoromass}
\rho \vec{v_{in}} = \curl \curl (W (\hat r)) ~~,
\end{equation}
where $\hat r$ is a unit vector in the radial direction.  We then use the method of separation of variables to express the radial mass flux as:

\begin{equation}\label{spvel}
\rho v_{inr}(r, \theta, \phi, t) = -\nabla_h^2 W = H(\theta, \phi, t) F(r) ~~.
\end{equation}

Substituting this into Equation (\ref{continuity}) and evaluating it at $r = R$ gives an expression for $H(\theta, \phi)$ in terms of the divergence of the surface flow in Equation (\ref{hvs})

\begin{equation}\label{Htp}
H(\theta, \phi, t) = - \frac{1}{\delta} \nabla_h \cdot \vec{v}_{ins}
\end{equation}

where
\begin{equation}
\delta = \left.\frac{1}{\rho r^2} \frac{\partial}{\partial r} \left(r^2 F(r)\right)\right|_{r=R} ~ .
\end{equation}

The radial profile $F(r)$ is defined as
\begin{equation}
F(r) = 4 x (1-x)  ~~ (r > r_p)
\end{equation}
where
\begin{equation}
x = \frac{r - r_p}{R-r_p}    
\end{equation}
and $r_p = 0.9 R$ is a specified penetration depth.  $F(r) = 0$ for $r \leq r_p$.

Applying a spherical harmonic transform to Equation (\ref{Htp}), indicated by a tilde, then yields 
\begin{equation}\label{Hlm}
\tilde{H}_{\ell m}(t) = - \frac{1}{\delta} \widetilde{\nabla_h \cdot \vec{v}_{ins}}
\end{equation}
 
 After doing some mathematical manipulation the 3D magnetic field dependent inflow velocities are explicitly defined as follows:

\begin{eqnarray}\label{inflowr}
 v_{in r}(r,\theta,\phi,t) = \frac{F(r)}{ \rho} \sum_{\ell, m}
 ~\left( \tilde{H}_{\ell m} Y_{\ell m}(\theta,\phi) \right) ~~~,
\end{eqnarray}
\begin{eqnarray}\label{inflowtheta}
 v_{in\theta}(r,\theta,\phi,t) =  - \frac{1}{\rho r\sin\theta} \frac{\partial}{\partial\theta}
 \sum_{\ell,m} \left( [2rF(r) + r^2F'(r)]\frac{1}{\ell (\ell + 1)} \tilde{H}_{\ell,m} Y_{\ell m}(\theta,\phi) \right) ~~~,
\end{eqnarray}
\begin{eqnarray}\label{inflowphi}
 v_{in\phi}(r,\theta,\phi,t) = - \frac{1}{\rho r\sin\theta} \frac{\partial}{\partial\phi}
 \sum_{\ell,m} \left( [2rF(r) + r^2F'(r)]\frac{1}{\ell (\ell + 1)} \tilde{H}_{\ell,m} Y_{\ell m}(\theta,\phi) \right) ~~~.
\end{eqnarray}

  \begin{figure}    
   \centerline{\hspace*{0.015\textwidth}
               \includegraphics[width=0.425\textwidth,clip=]{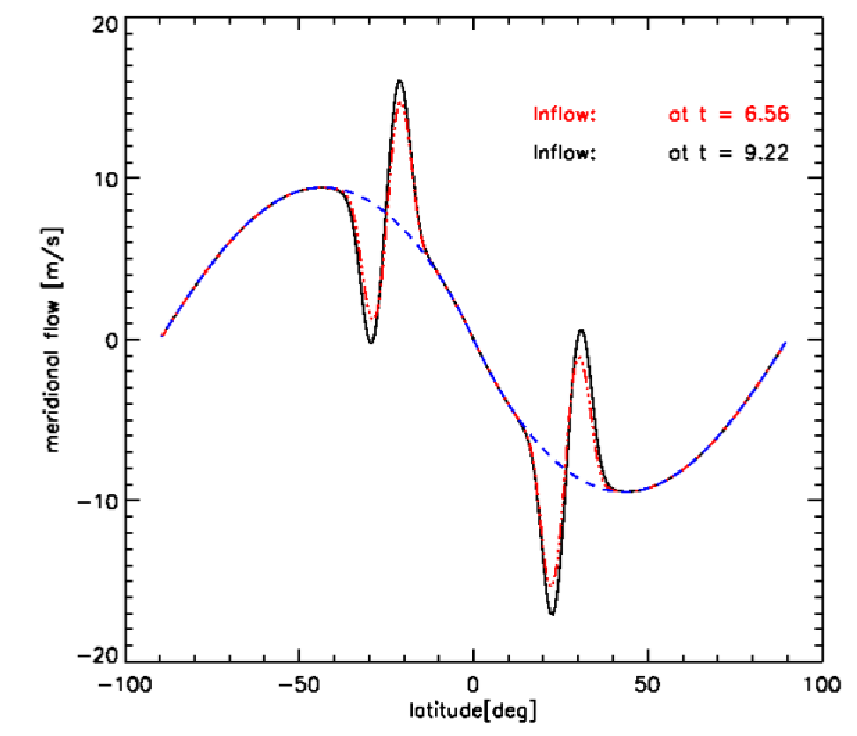}
               \hspace*{-.01\textwidth}
               \includegraphics[width=0.600\textwidth,clip=]{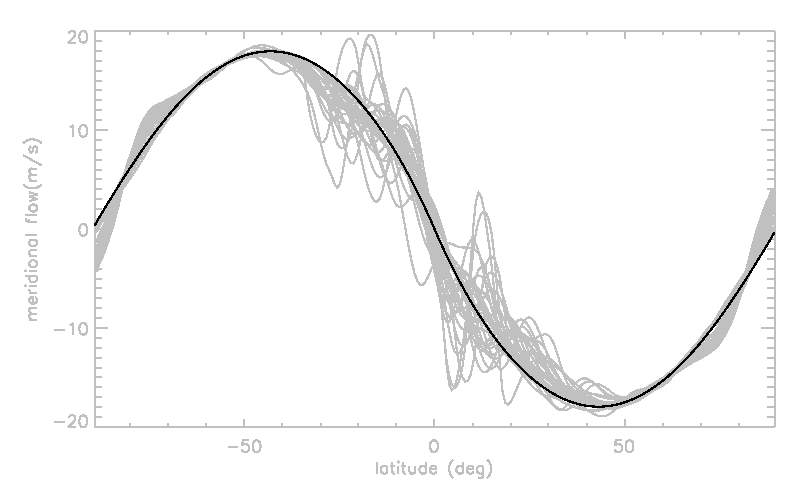}
              }
     \vspace{-0.35\textwidth}   
     \centerline{\Large \bf     

      \hspace{0.06 \textwidth}  \color{black}{($a$)}
      \hspace{0.625\textwidth}  \color{black}{(b)}
         \hfill}
     \vspace{0.35\textwidth}    
              
\caption{STABLE illustrates the effect of BMR inflows on the surface meridional flow ($v_\theta$). ($a$) A simulation with a single pair of sunspots in each hemisphere (see Figure \ref{F-mag-Br}). Positive flow velocities are directed southward. The blue curve shows the meridional flow profile without BMR inflows while the red and black curves show added BMR inflow speeds of different amplitudes (Equation \ref{inflowtheta}), as indicated. ($b$) Meridional flow variations (in grey) over the course of one magnetic cycle for a dynamo simulation (Case $C1$; Table 1), with many BMRs in each hemisphere.}
   \label{F-meridional-flow}
   \end{figure}

The resulting radial and longitudinal components are illustrated in Figures \ref{F-vr-inflow} and \ref{F-vphi-inflow}. From this we see that the converging flows toward the BMRs at the surface produce a downflow in the center surrounded by a narrow annular upflow.  The horizontal inflow near the surface becomes an outflow deeper down (Figure \ref{F-vphi-inflow}), which is consistent with helioseismic studies (see Figure 11 of \citet{hindman09}).

Thus, as described in Equation~\ref{eq:tflows}, the total flow field now available to STABLE includes not just the differential rotation and meridional circulation, but also the 3D BMR inflow components. In particular, the flow along the heliographic longitudinal direction is the summation of the longitudinal component of the inflow velocity (Equation~\ref{inflowphi}) and the axisymmetric differential rotation profile. Similarly, the complete flow field in the radial and latitudinal directions incorporate those components of the inflow velocity (Equation~\ref{inflowr}~and ~\ref{inflowtheta}).

This has implications for solar cycle variation. An observational study by \citet{hath14} reported that the inflow around the sunspot zones superimposed on the poleward meridional flow profile created a weakening of the meridional flow on the poleward sides of the active (sunspot) latitudes. Another study by \citet{braun19} reported the contribution of converging inflows to longitudinal averages of zonal and meridional flows had an influence on modifying global flows.

Thus, BMR inflows may contribute to the solar cycle's global variation in this meridional flow in a manner that could help explain the variations in the strength and duration of solar activity cycles \citep{camers10, camer12}. Indeed, the fully 3D STABLE model shows that the effect of the latitudinal BMR inflows ($v_{in\theta}$; Equation \ref{inflowtheta}) is to alter the mean meridional circulation that transports poloidal field toward the poles at the surface (see Figure \ref{F-meridional-flow}). 

\section{Impacts of the BMR inflows on the Polar Field Buildup }
\label{sec:inflowpolarfield}

The polar field at the given solar activity minimum reflects the source for the generation of the toroidal magnetic field for the subsequent cycle by the mechanism of differential rotation \citep{camer12, chrb20}. In order to study its buildup, we begin by conducting parameter studies of two BMRs (i.e., a pair of sunspots in both hemispheres as shown in Figure \ref{F-mag-Br}). In particular, we vary the magnitude of the BMR inflows and

degree of smoothing applied to the BMR unsigned magnetic field (Equation~\ref{hvs}). 
In this way, we establish how individual BMRs and BMR inflows contribute to the build up of the polar field and polar flux.

\begin{figure}  
   \centerline{\includegraphics[width=1.0\textwidth,clip=]{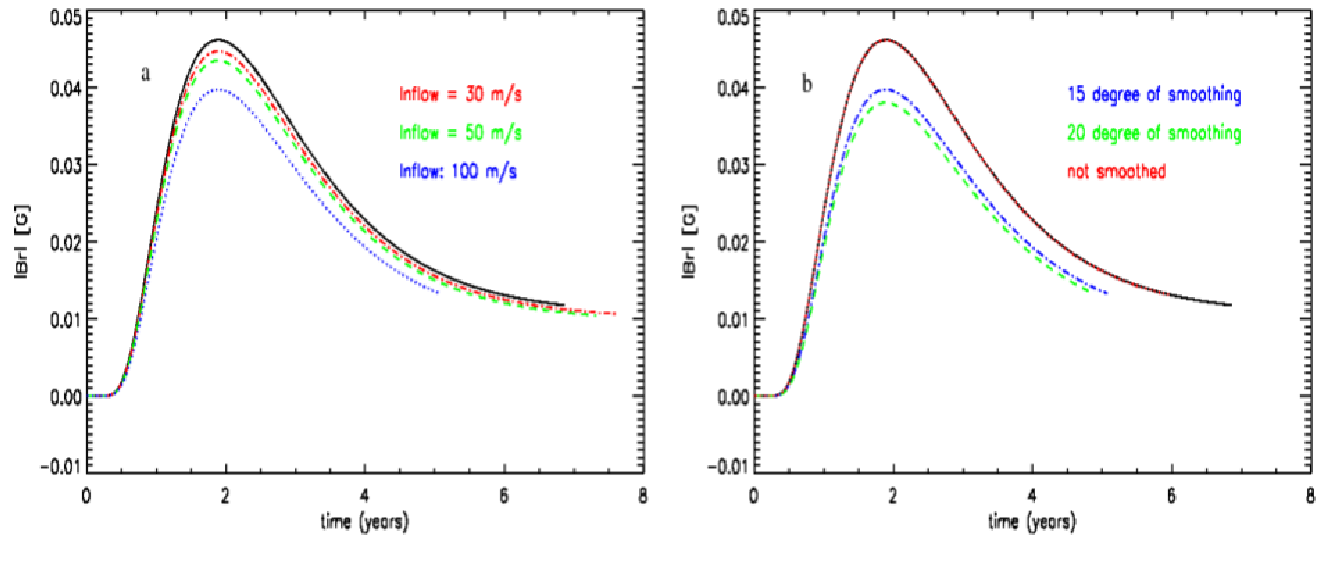}
              }
\caption{The time evolution of the polar magnetic field strength $B_r$ (average radial field poleward of $\pm 60^{\circ}$  latitude) as a function of time for different parameter studies 
for two BMRs, one placed in each hemisphere at emergence angle latitudes $\pm 25^{\circ}$. $(a)$ the polar magnetic field buildup shown for different choices of the inflow velocity: $v_{in}$ = 0 (black/solid curve), $v_{in}$ = 30 m $s^{-1}$ (red/dotted curve), $v_{in}$ = 50 m $s^{-1}$ (green/dashed curve), and $v_{in}$  = 100 m $s^{-1}$ (blue/dash-dot curve), for a smoothing radius of ($15^{\circ}$).  
$(b)$ polar buildup demonstrating dependence on the area smoothing: no inflow (black/solid line), no smoothing (red/dotted line),  smoothing radius $15^{\circ}$ (blue dot-dashed), $20^\circ$ (green/dashed). 
The magnetic field is in Gauss and time is given in years.}
\label{F-polarfield}
   \end{figure}  
   
As Figure \ref{F-mag-Br} showed, 
BMRs are distorted by differential rotation, diffused, and advected poleward, resulting in a cancellation of leading flux at the equator and an accumulation of trailing flux at the poles.   This accumulation of polar flux is suppressed by the presence of inflows, as demonstrated in Figure \ref{F-polarfield}.

We find that the converging flows enhance flux cancellation in BMRs by bringing together the opposite magnetic polarities.   Stronger inflows lead to more flux cancellation, which suppresses polar field generation as shown in Fig. \ref{F-polarfield}$a$.  

The polar field buildup also depends on the degree of smoothing (Figure  \ref{F-polarfield}$b$), with greater smoothing leading to more effective flux cancellation
and so a reduction in the amplitude of the BMR magnetic field and subsequently in the polar field buildup.
An increase of the smoothing area from 15$^\circ$ to 20$^\circ$ leads to a slight reduction in the polar field generation.  
Finally, we note that, as in previous studies that do not include BMR inflows \citep[e.g.]{hazra16}, STABLE with inflows finds that placing pairs of BMRs at higher latitudes are more effective at generating polar field, and that there is also a dependence on the separation distance between the leading and trailing polarities of the BMR \citep{miesc16}, such that larger spacing leads to less flux cancellation and stronger polar fields.

\section{Impacts of BMR Inflows on the STABLE Solar Dynamo and Magnetic Cycles }
\label{sec:solaractivity}

\begin{table}

\begin{tabular}{lcccccccc} 
  \hline
Case & ${\Phi_0}$ & $\eta_{CZ}$ & $\gamma_{CZ}, \gamma_S$& $c$ & $B_{tor} ~ ~B_{pol} $ &  Period & $\#$ of \\
     &  (G cm$^2$)  &(cm$^2$ s$^{-1}$)      & (cm s$^{-1}$)& m/s & (kG) ~ (G)& (yr) & (reversals) \\
  \hline
T & 2.4 & $1.5 \times 10^{12}$  & 2, 20 & -- & 44 ~~2160 & 10.6 & 14  \\

\hline
C1 & 2.4 & $1.5 \times 10^{12}$ & 2, 20 & 10 & 19 ~~686 & 12.3 & 11 \\ 
C2 & 2.4 &  $1.5 \times 10^{12}$ & 2, 20 & 30  & -- & -- & -- & \\ 
  \hline
\end{tabular}
\caption{STABLE dynamo simulations. See text for an explanation of cases (Column 1). $\Phi_0$ (Column 2) is the BMR magnetic field strength parameter (Equation 8,  \citep{karak17}). $\eta_{CZ}$ (Column 3) is the turbulent diffusion evaluated at the middle of the convection zone; $\gamma_{CZ}$ is the turbulent pumping evaluated at the middle of the convection zone, and $\gamma_{S}$ is the same at the surface (Column 4) -- see Equations 3-4 of \citet{karak17}. $c$ (Column 5) is the inflow parameter with units of velocity (Equation~\ref{hvs}).  See text for discussion of simulation output, including rms magnetic field values (Columns 6-7) and cycle periods and number of reversals (Columns 8-9).}
\label{T-inflow}
\end{table}

Table \ref{T-inflow} is  a summary of the STABLE dynamo simulations presented in this paper.  The T case is a simulation from the tilt-angle-quenched dynamo (without BMR inflows) that uses the same STABLE parameters as of Case B9 discussed in \citet{karak17}. Both of the C simulations include magnetic-field-dependent BMR inflows as described in this paper but do not have tilt-angle quenching turned on, as discussed above. Root-mean-square (rms) values listed in Columns 5-6 are simulation output and are based on integrals over the entire computational volume and quoted for the mean toroidal field ($B_{tor}$) and the mean poloidal field ($B_{pol}$). All simulation runs described in this paper have spatial resolutions of 200 $\times$ 256 $\times$ 512 in r, $\theta$, and $\phi$, respectively. 

As in previous works of STABLE by \citet{miesc14}, \citet{miesc16}, \citet{hazra16}, \citet{karak17}, the Babcock-Leighton mechanism in these simulations represents the poloidal field source and is crucial for the buildup and reversals of the polar fields (see also section \ref{sec:inflowpolarfield}). The evolution of the polar field is a consequence of Hale's polarity rules, together with the systematic tilt angle distribution as observed and described by Joy's law. 

\begin{figure}    
   \centerline{\includegraphics[width=1.0\textwidth,clip=]{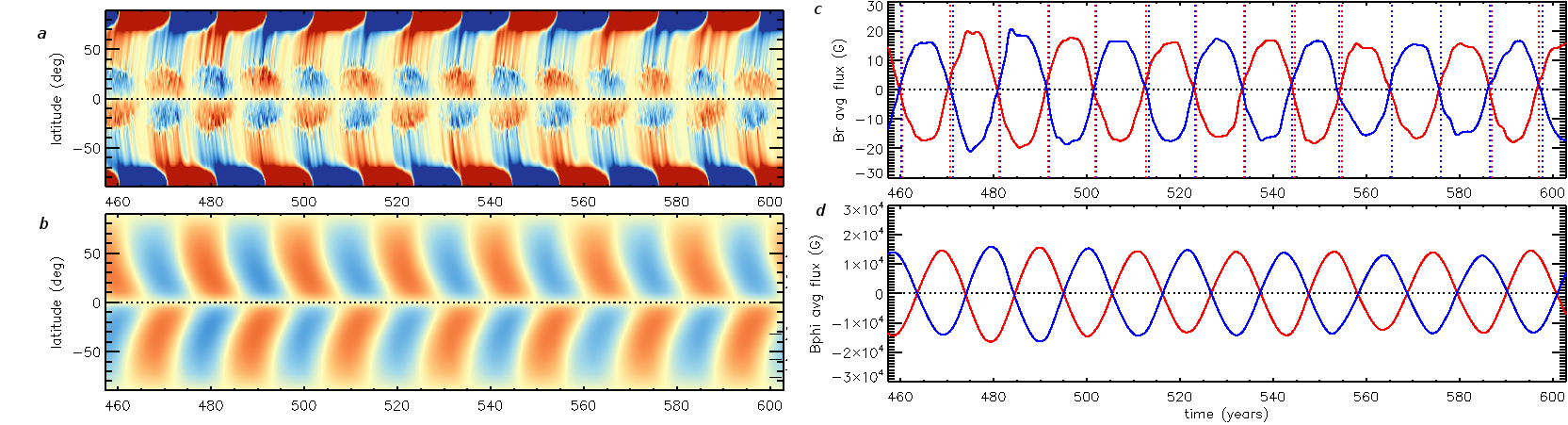}
              }
\caption{Magnetic cycles in Case T. ($a$) Butterfly diagram of the longitudinally-averaged radial field $\left<B_r\right>$ at the surface ($r = R$) as a function of latitude and time, highlighting fourteen magnetic cycles. Peak amplitudes can exceed 300 G but the color table saturates at $\pm$ 100G.
 ($b$) Diagram of longitudinally-averaged toroidal field $\left<B_\phi\right>$ near the base of the convection zone ($r = 0.72 R$). Red and blue denote eastward and westward field respectively. 
 ($c$) $\left<B_r\right>$ averaged over the north (blue) and south (red) polar regions, above latitudes of $\pm$ 70$^\circ$, plotted vs. time. Vertical dotted lines mark polar field reversals in the NH (blue) and SH (red). ($d$) is similar to ($c$) but for $\left<B_\phi\right>$ in the lower convection zone and averaged over the entire NH (blue) and SH (red), as opposed to just the polar regions as in ($c$), with a saturation level for the color table of 90 kG.}
   \label{F-tilt-que}
   \end{figure}

\begin{figure}[h]    
   \centerline{\includegraphics[width=1.0\textwidth,clip=]{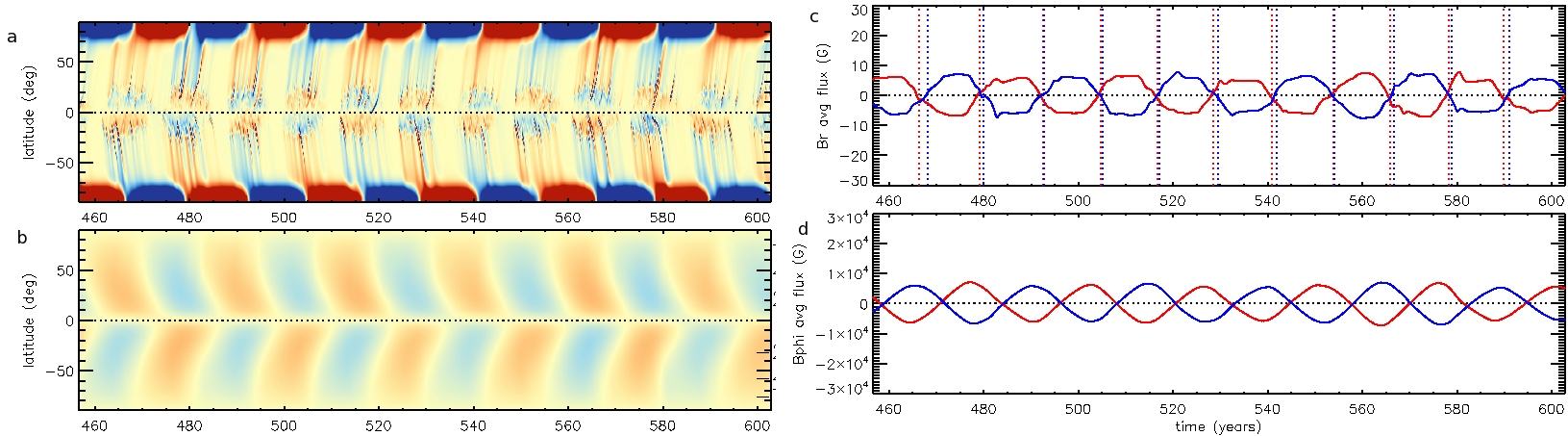}
              }
\caption{Magnetic cycles in Case C1, with BMR-inflow saturation. The quantities shown, color table saturation levels (a,b), and axis ranges (c,d) are the same as in Figure~\ref{F-tilt-que}.}
   \label{F-inflow-que}
   \end{figure}

\begin{figure}
   \centerline{\includegraphics[width=0.6\textwidth,clip=]{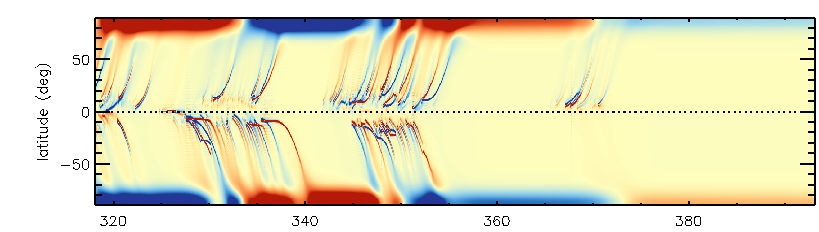}
              }
              \caption{Magnetic cycles in Case C2, BMR-inflow case,  displayed as in Figure~\ref{F-tilt-que}. A larger inflow amplitude of 30 m/s leads to a shutdown of the cycling dynamo.}
   \label{F-shutdown}
   \end{figure}

The simulations summarized in Table \ref{T-inflow} correspond to a diffusion-dominated flux-transport dynamo model with enough magnetic flux in the BMRs to achieve sustained dynamo action. 
Similar to the previous works, we determine the mean radial surface magnetic field, $\left<B_r\right>$, and the mean toroidal field at the base of the convection zone, $\left<B_{\Phi}\right>$, and study 
the evolution of global magnetic field over many magnetic solar cycles, including solar cycle modulation and average period and number of reversals over about 140 years (Columns 8-9 in Table~\ref{T-inflow}), with results for cases T, C1, and C2 shown in Figure~\ref{F-tilt-que}-\ref{F-shutdown}.

Case $T$ is reproduced using the parameters from \citet{karak17} Case B9 and relies on tilt-angle quenching to saturate the dynamo, with no BMR inflows. Its evolution is shown in Figure \ref{F-tilt-que} over many magnetic cycles. The buildup of a polar magnetic field in Figure \ref{F-tilt-que} $c$ naturally arises from the tilt angles of emergent BMRs combined with differential rotation, turbulent diffusion and poleward meridional flow. Because of the tilt-angle quenching described in Equation~\ref{tiltangle_sat}, the dynamo is saturated and a sustained cycle is achieved. Figure \ref{F-tilt-que} ($a$) shows this through a butterfly diagram of the longitude-averaged $\left<B_r\right>$ in a time-latitude plot. One clearly sees the emergence of sunspots (BMRs) at progressively lower latitudes and also the poleward advection of the magnetic field by the meridional circulation at higher latitudes, reversing the polar fields \citep{karak17}.  

The BMR-inflow Cases $C1$ and $C2$ use only Joy's Law tilts with no tilt-angle quenching term. The smoothing is done using a circular tent filter of radius $7.5^{\circ}$. The calibration factor $B_n$ in Equation (\ref{hvs}) is set to 1480 G, and the factor $c$ in that equation is set for the Cases $C1$ and $C2$ as described in Table~\ref{T-inflow}. Other parameters are as in Case $T$. 

Figure~\ref{F-inflow-que} shows that, as in Case $T$, a cycling dynamo is sustained for Case $C1$ as the distortion and dispersal of these tilted (Joy's law) BMRs by differential rotation, meridional circulation, BMR inflows, and turbulent diffusion give rise to a poleward migration of trailing flux that reverses the polar fields as described in the previous STABLE models \citep{miesc14, miesc16, hazra16} and \citep{karak17}. However, in Case $C1$ we see a clear difference; the magnetic flux in the BMRs is very weak compared to Case $T$. This is because the converging flows into the BMRs enhance flux cancellation as discussed in Section~\ref{sec:inflowpolarfield}, affecting the polar field buildup. As seen in Figure~\ref{F-inflow-que}$c$, this polar field buildup is reduced compared to Figure~\ref{F-tilt-que}$c$. 

\begin{figure}    
   \centerline{\hspace*{0.015\textwidth}
               \includegraphics[width=0.515\textwidth,clip=]{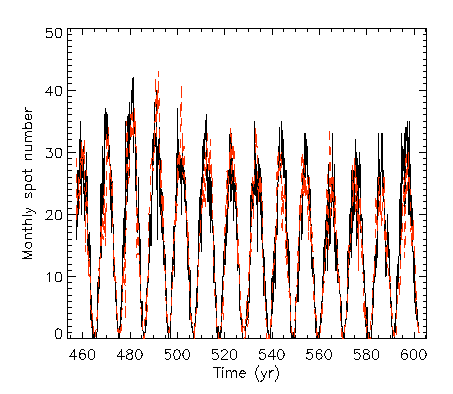}
               \hspace*{-0.03\textwidth}
               \includegraphics[width=0.515\textwidth,clip=]{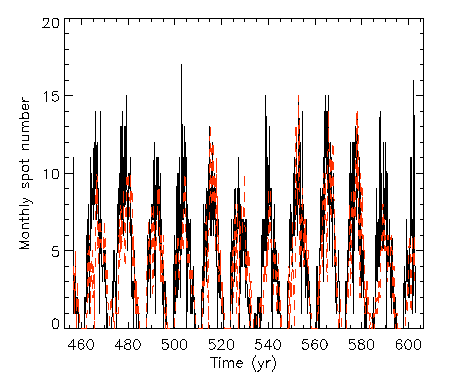}
              }
     \vspace{-0.40\textwidth}   
     \centerline{\Large \bf     
      \hspace{0.07 \textwidth}  \color{black}{($a$)}
      \hspace{0.45\textwidth}  \color{black}{(b)}
         \hfill}
     \vspace{0.36\textwidth}    
              
\caption{Sunspot (BMR) number in ($a$) Case T and ($b$) Case C1, averaged over 1-month intervals, for the time span shown in Figures \ref{F-tilt-que} and \ref{F-inflow-que}. Black and red represent the northern and southern hemispheres respectively.}
   \label{F-sunspot-number}
   \end{figure}

This reduced polar field buildup produces weaker cycles since the polar field is a measure of the poloidal field and provides the seed for the toroidal field in the subsequent cycle (see Figure \ref{F-inflow-que}$b$ and $d$).
Figure \ref{F-sunspot-number} illustrates the sunspot number (BMR number) vs time as produced by case $T$ ($a$) and Case $C1$ ($b$), corresponding to Figures~\ref{F-tilt-que} and \ref{F-inflow-que}, respectively, and shows that the weaker cycle amplitudes in Case $C1$ is reflected by a reduction in the number of BMRs.  This arises because the weaker poloidal fields have led to weaker toroidal fields (see the amplitudes of Figure \ref{F-inflow-que}$d$ vs \ref{F-tilt-que}$d$), which control the number of BMRs (see Equation~\ref{eq:spotoma}).
The BMR emergence in turn controls the surface flux that produces the poloidal field.
This is the nonlinear feedback mechanism at the heart of the Babcock-Leighton paradigm that explains why the polar field at  a given solar cycle minimum is generally correlated with the sunspot number of the next solar cycle \citep{munoz13}.  
   
Moreover, in the framework of STABLE, the time delay distribution of BMR emergence is dependent on the magnetic field through a lognormal distribution with mean and mode ($\tau_s$ and $\tau_p$) as defined by Equation~\ref{eqch5:mean}. If the toroidal field at the base of the convection zone is weaker, it increases $\tau_s$ and $\tau_p$, making the BMR emergence less frequent. This makes the poloidal field production slower, which ultimately causes weaker fields, slower polarity reversals, and longer cycles.   The average cycle period of Case $C1$ is longer (12.3 years) compared to Case $T$ (10.5 years) (see Table \ref{T-inflow}). Consequently, Figure \ref{F-tilt-que} covers 14 magnetic cycles in 140 years for Case $T$, while Figure \ref{F-inflow-que} shows only 11 cycles in the same time interval for Case $C1$.
 
Another factor that contributes to the amplitude and timing of cycles is the meridional circulation, which transports the poloidal field to the poles and which is affected by the introduction of BMR inflows ( see Figure \ref{F-meridional-flow} $b$). Thus, the magnetic field dependent BMR inflows modulate the solar cycle and affect the cycle periods in more than one way.

\begin{figure}    
   \centerline{\includegraphics[width=1.0\textwidth,clip=]{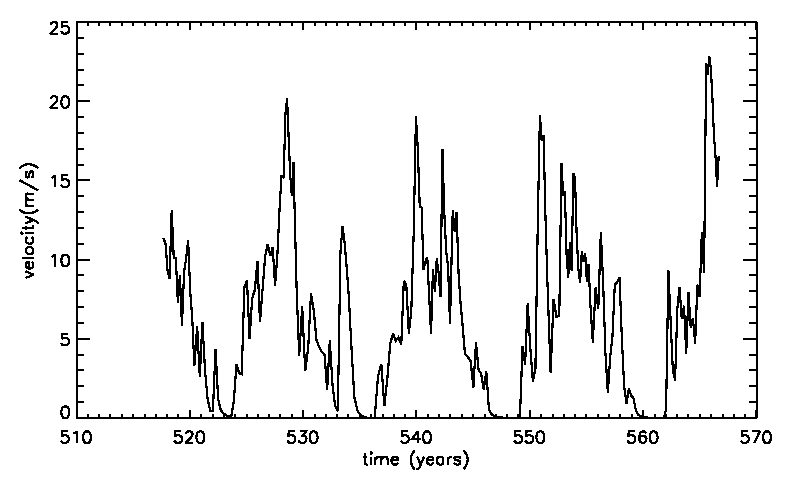}          }
\caption{The maximum inflow velocity at the surface for all BMRs at a given time is shown for Case $C1$ over about four magnetic cycles.  Here the control parameter $c$ is set to 10 m s$^{-1}$.  The corresponding butterfly diagram is shown in Figure \ref{F-inflow-que}$a$.}
   \label{F-inflow-vrms}
   \end{figure}

As discussed in Section \ref{sec-con-flow}, the mean amplitude of the BMR inflows is regulated by the control parameter $c$ and the calibration factor $B_n$.  However, as Equation~\ref{hvs} indicates, the inflow velocity distribution at any one time also depends on the gradient of the smoothed surface field $B_r$, which varies over the course of a magnetic cycle.  This is illustrated in Figure \ref{F-inflow-vrms} which shows that the maximum inflow speed in Case $C1$ varies from zero, when there are no BMRs, to about 20 m s$^{-1}$ at cycle maximum.  This is well within the range of 20-50 m/s inferred from helioseismology (Section~\ref{sec-con-flow}).  In other words, the observed amplitude of BMR inflows is more than sufficient to act as a dynamo saturation mechanism in our model.

Finally, as summarized in Table \ref{T-inflow}, the two Case $C$ simulations were run with two different magnitudes of the BMR-inflow amplitude parameter $c$. Our simulations show that the larger magnitude of converging flows into the BMRs noticeably affect the evolution of the surface magnetic fields and the amplitude of magnetic cycles. Indeed, the Case $C2$ choice of $c=30 m/s$ shuts down the dynamo as demonstrated in Figure \ref{F-shutdown}. This Case was initialized from Case $C1$, but after several cycles, the enhanced cancellation of magnetic flux in the BMRs reduced the field amplitudes to the point where they become too weak to initiate new BMRs, leading to the eventual and irreversible decay of all fields.

\section{Conclusion} 
      \label{S-Conclusion} 

We have demonstrated that BMR inflows are likely to have a substantial impact on the solar dynamo.
Converging flows towards active regions are an observed physical phenomenon and should not be neglected. With the work reported here, the STABLE dynamo model has been enriched to include this new feature and we study the effect of the converging flows towards the BMRs on the evolution of the large-scale magnetic field, dynamo saturation, and modulation of magnetic cycles.  To our knowledge, this is the first fully-3D Babcock-Leighton dynamo model to include BMR inflows and reproduce prominent features of the solar cycle.

Our simulations demonstrate that BMR inflows can operate as a viable dynamo saturation mechanism.  The inflows depend on the surface distribution of magnetic flux in BMRs and therefore provide a nonlinear feedback that can prevent unlimited growth of the magnetic energy.
We determine the converging inflow velocity from the surface magnetic field, $B_r$, which includes the explicit emergence and dispersal of tilted BMRs under the influence of differential rotation, turbulent diffusion, and meridional circulation.  The induced inflows feed back on $B_r$ by enhancing magnetic flux cancellation, altering the evolution and inhibiting the Babcock-Leighton mechanism.  Consequently, for the first time, stable dynamo solutions are achieved in which the inflows alone regulate the amplitudes and periods of magnetic cycles.

We conduct parameter studies to identify the factors that most influence the efficiency of poloidal field production from the dispersal of BMRs. These parameter studies are conducted first for one BMR in the northern hemisphere and then for two BMRs, one in each solar hemisphere. We find the inflow reduces the buildup of the polar flux because larger inflow speeds enhance flux cancellation and inhibit poloidal field generation.  The efficiency of poloidal field generation also depends on the separation between BMR polarities, smoothing, and latitude of emergence (Figure~\ref{F-polarfield}).  This parameter study can also help us to understand the effect of the converging flows on the evolution of individual BMRs. Indeed, we find that BMR inflows can impact both the global poloidal field through enhanced local flux cancellation, and the nature of the local flux evolution within each BMR.

This is borne out by our cycling dynamo simulations that incorporate multiple BMRs and indicate that the reduced efficiency of poloidal field generation due to magnetic field dependent BMR inflows has a major influence on the structure of the global magnetic field and the nature of magnetic cycles.  We find a reduced number of BMRs in the inflow-saturated dynamo (Case $C1$) compared to the tilt-angle saturated dynamo (Case $T$). This is associated with reduced poloidal and toroidal magnetic fields due to the inflow-driven flux cancellation within the BMRs.  The cycle period of polarity reversals also changes significantly, increasing from 10.6 years in Case $T$ to 12.3 years in Case $C1$.  This is mainly attributed to an inhibition of the polar flux buildup for the BMR-inflow case.  BMR inflows also modify the surface meridional flow (Figure~\ref{F-meridional-flow}). 

When the amplitude of the BMR inflow is increased from approximately 10 m s$^{-1}$ to 30 m s$^{-1}$, the dynamo shuts down.   Thus, it is clear that the inflow amplitudes of 10-50 m s$^{-1}$ inferred from helioseismic observations (Section~\ref{sec-con-flow}) are more than sufficient to have a substantial impact on the operation of Babcock-Leighton solar dynamo models, and, as we have shown, can indeed saturate the dynamo in a manner resulting in solar-like cycles.  The precise thresholds at which the dynamo shuts down vs. stable cycles vs. becomes unbounded likely depends on the details of the model configuration.

In conclusion, for the first time, we have presented a self-excited 3D kinematic solar dynamo model solution with solar-like cycles that is totally sustained by the observed distribution of tilted BMRs and observed magnitudes of BMR inflows. This is important because an accurate representation of the solar dynamo is 
needed for accurate predictions of the solar cycle and related space-weather activity, and to better understand the fundamental nature of the evolution of the magnetic field in the Sun.
\begin{acks}
The authors thank the the National Center for Atmospheric Research (NCAR) Advanced Study Program and the HAO Visitor Program for funding KT's visit to HAO/NCAR to support this research. The computations were performed using resources provided by NCAR’s Cheyenne. NCAR is a major facility sponsored by the NSF under Cooperative Agreement No. 1852977.
\end{acks}

\end{article} 

\end{document}